\documentclass[aps,prl%
,twocolumn%
,showpacs%
,superscriptaddress%
]{revtex4}

\usepackage{graphicx}
\usepackage{times}

\begin{document}

\newcommand{\be}{\begin{equation}}
\newcommand{\ee}{\end{equation}}
\newcommand{\bea}{\begin{eqnarray}}
\newcommand{\eea}{\end{eqnarray}}
\newcommand{\f}{\frac}
\newcommand{\p}{\partial}
\newcommand{\no}{\nonumber}
\newcommand{\kT}{k_{\rm B}T}
\newcommand{\e}{{\rm e}}
\newcommand{\dd}{{\rm d}}

\newcommand{\eg}{{\it e.g.}}
\newcommand{\ie}{{\it i.e.}}
\newcommand{\pc}{p_{\rm c}}

\newcommand{\kav}{\left<k\right>}

\newcommand{\affila}{
 Department of Biological Physics, E\"otv\"os University,
 P\'azm\'any P.\ stny.\ 1A, H-1117 Budapest, Hungary
}
\newcommand{\affilb}{
 Biological Physics Research Group of HAS,
 P\'azm\'any P.\ stny.\ 1A, H-1117 Budapest, Hungary
}

\title{
 Clique percolation in random networks
}

\author{Imre Der\'enyi}
 \affiliation{\affila}
\author{Gergely Palla}
 \affiliation{\affilb}
\author{Tam\'as Vicsek}
 \affiliation{\affila}
 \affiliation{\affilb}

\date[]{\lowercase{s}ubmitted to Phys.\ Rev.\ Lett.\ 10 November 2004;
        accepted for publication 20 April 2005}

\begin{abstract}
The notion of $k$-clique percolation in random graphs is introduced,
where $k$ is the size of the complete subgraphs whose large scale
organizations are analytically and numerically investigated. For the
Erd\H{o}s-R\'enyi graph of $N$ vertices we obtain that the percolation
transition of $k$-cliques takes place when the probability of two
vertices being connected by an edge reaches the threshold
$\pc(k)=[(k-1)N]^{-1/(k-1)}$. At the transition point the scaling
of the giant component with $N$ is highly non-trivial and depends
on $k$. We discuss why clique percolation is a novel and efficient
approach to the identification of overlapping communities in large
real networks.
\end{abstract}



\maketitle

There has been a quickly growing interest in networks, since
they can represent the structure of a wide
class of complex systems occurring from the level of cells to
society. Data obtained on real networks show that the
corresponding graphs exhibit unexpected non-trivial properties,
\eg, anomalous degree distributions, diameter,
spreading phenomena, clustering coefficient, and correlations
\cite{watts-strogatz,barabasi-albert,albert-revmod,dm-book,barrat}.
Very recently great attention has been paid to the local structural
units of networks. Small and well defined subgraphs
have been introduced as ``motifs''
\cite{alon}.
Their distribution and clustering properties
\cite{alon,vazquez-condmat,kertesz-condmat}
can be used to interpret global features as well.
Somewhat larger units, made up of vertices that are more densely
connected to each other than to the rest of the network, are often
referred to as communities
\cite{domany-prl,gn-pnas,zhou,newman-fast,al-parisi,huberman,spektral,potts},
and have been considered to be the essential structural units of real
networks.
They have no obvious definition, and most of the recent methods for their
identification rely on dividing the network into smaller
pieces. The biggest drawback of these methods is that they do not allow
for overlapping communities, although overlaps are generally assumed to
be crucial features of communities.
In this Letter we lay down the fundamentals of a kind of percolation
phenomenon on graphs, which can also be used as an effective and
deterministic method for uniquely identifying overlapping communities
in large real networks
\cite{Pallasubm}.

Meanwhile, the various aspects of the classical Erd\H{o}s-R\'enyi
(ER) uncorrelated random graph
\cite{e-r}
remain still of great interest
since such a graph can serve both
as a test bed for checking all sorts of new ideas concerning
complex networks in general, and
as a prototype to which all other random graphs can be
compared.
Perhaps the most conspicuous early
result on the ER graphs was related to the percolation transition
taking place at $p=\pc \equiv 1/N$, where $p$ is the probability that
two vertices are connected by an edge and $N$ is the total number
of vertices in the graph. The appearance of a {\em giant component},
which is also referred to as the {\em percolating component},
results in a dramatic change in the overall topological features
of the graph and has been in the center of interest for
other networks as well.

In this Letter we address the general question of subgraph
percolation in the ER model. We obtain analytic
and simulation results related to the appearance of a giant
component made of complete subgraphs of $k$ vertices
($k$-cliques).
In particular, we provide an analytic expression for the
threshold probability at which the percolation transition of
$k$-cliques takes place.
The transition is continuous, characterized by
non-universal critical exponents, which depend on both $k$ and the way
the size of the giant component is measured. Our analytic calculations
are in full agreement with the corresponding numerical
simulations.

\begin{figure}[b!]
\centerline{\includegraphics[angle=0,width=0.85\columnwidth]{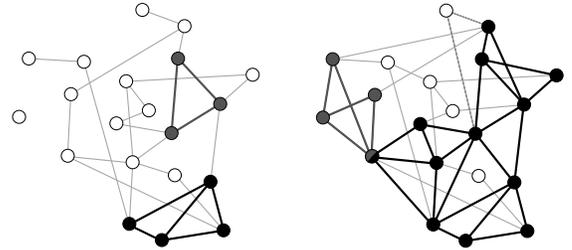}}
\caption{
Sketches of two ER graphs of $N=20$ vertices and with edge
probabilities $p=0.13$ (left one) and $p=0.22$ (right one, generated by
adding more random edges to the left one). In both cases all the edges
belong to a ``giant'' connected component, because the edge
probabilities are much larger than the threshold ($\pc \equiv 1/N =
0.05$) for the classical ER percolation transition. However, in the
left one $p$ is below the 3-clique (triangle) percolation threshold, $\pc(3)
\approx 0.16$, calculated from Eq.\ (\ref{pcrit}), therefore, only two
small 3-clique percolation
clusters (distinguished by black and dark gray edges) can be observed.
In the right graph, on the other hand, $p$ is above this threshold and,
as a consequence, most 3-cliques accumulate in a ``giant'' 3-clique
percolation cluster (black edges). This graph also exhibits an
overlap (half black, half dark gray vertex) between two 3-clique
percolation clusters (black and dark gray).
}
\label{fig_sketch}
\end{figure}

Before we proceed to calculate the threshold and the exponents we need
to outline some basic definitions. {\em $k$-cliques}, the central
objects of our investigation, are defined as complete (fully connected)
subgraphs of $k$ vertices
\cite{bollobas-book}.
As an illustration, in Fig.\ \ref{fig_sketch} all the 3-cliques
(triangles) are emphasized with either black or dark gray edges.
We also introduce a few new notions specific to our problem.
(i) {\em $k$-clique adjacency}: two $k$-cliques are adjacent if they
share $k-1$ vertices, \ie, if they differ only in a single vertex.
(ii) {\em $k$-clique chain}: a subgraph, which
is the union
of a sequence of adjacent $k$-cliques.
(iii) {\em $k$-clique connectedness}: two $k$-cliques are
$k$-clique-connected if they are parts of a $k$-clique chain.
(iv) {\em $k$-clique percolation cluster (or component)}: it is a
maximal $k$-clique-connected subgraph, \ie, it
is the union
of all $k$-cliques that are $k$-clique-connected to a particular
$k$-clique. This is illustrated in Fig.\ \ref{fig_sketch}, where both
graphs contain two 3-clique percolation clusters, the smaller ones in
dark gray and the larger ones in black.
We note that these objects can be considered as interesting
specific cases of the general graph theoretic objects defined
in Refs.\ 
\cite{clique-overlap}
and
\cite{short-cycles}
in very different contexts.

A $k$-clique percolation cluster is very much like a regular (edge)
percolation cluster in the
{\em $k$-clique adjacency graph}, where the vertices represent the
$k$-cliques of the original graph, and there is an edge between two
vertices if the corresponding $k$-cliques are adjacent. Moving a
particle from one vertex of this adjacency graph to another one along
an edge is equivalent to {\em rolling} a {\em $k$-clique template} from one
$k$-clique of the original graph to an adjacent one. A $k$-clique template
can be thought of as an object that is isomorphic to a complete graph of
$k$ vertices. Such a template can be placed onto any $k$-clique of the
original graph, and rolled to an adjacent $k$-clique by relocating one
of its vertices and keeping its other $k-1$ vertices fixed. Thus, the
$k$-clique percolation clusters of a graph are all those subgraphs that
can be fully explored but cannot be left by rolling a $k$-clique template
in them.

Now, we present a general result for the threshold probability
(critical point) of $k$-clique percolation using heuristic arguments.
We find that a giant $k$-clique component appears in an ER graph
(as illustrated for $k=3$ in Fig.\ \ref{fig_sketch}) at
$p=\pc(k)$, where
\be
\pc(k)={1\over {[(k-1)N]^{1\over {k-1}}}}
.
\label{pcrit}
\ee
Obviously, for $k=2$ this result agrees with the known percolation
threshold ($\pc=1/N$) for ER graphs, because 2-clique connectedness is
equivalent to regular (edge) connectedness. Expression (\ref{pcrit})
can be obtained by requiring that after rolling a $k$-clique template from
a $k$-clique to an adjacent one (by relocating one of its vertices), the
expectation value of the number of adjacent $k$-cliques, where the
template can roll further (by relocating another of its vertices), be equal
to 1 at the percolation threshold.
The intuitive argument behind this criterion is that
a smaller expectation value would result in premature $k$-clique percolation
clusters, because starting from any $k$-clique the rolling would
quickly come to a halt and, as a consequence, the size of the clusters
would decay exponentially.
A larger expectation value, on the other hand, would allow an infinite
series of bifurcations for the rolling, ensuring that a giant cluster
is present in the system.
The above expectation value can be estimated
as $(k-1)(N-k-1)p^{k-1}$, where the first term $(k-1)$ counts the
number of vertices of the template that can be selected for the next
relocation, the second term $(N-k-1)$ counts the number of potential
destinations for this relocation, out of which only the fraction
$p^{k-1}$ is acceptable, because each of the new $k-1$ edges
(associated with the relocation) must exist in order to obtain a new
$k$-clique. For large $N$, our criterion can thus be written as
$(k-1)N\pc^{k-1}=1$, from which we get expression (\ref{pcrit}) for the
threshold probability. The above heuristic approach is similar
in spirit to the one used in Ref.\
\cite{cohen}
in the context of standard percolation on networks.

It is important to point out that {\em this result can be made
stronger} by a more detailed derivation which we shall present
elsewhere due to space limitations. In short, starting from the
distribution of the number of $k$-cliques adjacent to a randomly
selected one, and applying the so-called generating function formalism
\cite{newman-arbitPk},
one can derive the generating function of the distribution of the
number of $k$-cliques that can be visited from a randomly selected one.
This function diverges as $p$ approaches $\pc(k)$ from below, signaling
the threshold for percolation.
Furthermore, our result for $\pc(k)$ is also in perfect agreement with
the numerical simulations (see below).

There are two plausible choices to measure the size of the largest
$k$-clique percolation cluster. The most natural one, which we denote by
$N^*$, is the number of vertices belonging to this cluster. We can also
define an {\em order parameter} associated with this choice as the
relative size of that cluster:
\be
\Phi=N^*/N
. 
\ee
The other choice is the number ${\cal N}^*$ of $k$-cliques
of the largest $k$-clique percolation cluster (or equivalently, the number of
vertices of the largest component in the $k$-clique adjacency graph).
The associated order parameter is again the relative size of this
cluster:
\be
\Psi={\cal N}^*/{\cal N}
,
\ee
where ${\cal N}$ denotes the total number of $k$-cliques in the graph
(or the total number of vertices in the adjacency graph).
${\cal N}$ can be estimated as
\be
{\cal N} \approx
 {N \choose k}p^{k(k-1)/2}
\approx
 \f{N^k}{k!}p^{k(k-1)/2}
,
\label{Nk}
\ee
because $k$ different vertices can be selected in ${N \choose k}$
different ways, and any such selection makes a $k$-clique only if all
the $k(k-1)/2$ edges between these $k$ vertices exist, each with
probability $p$.
Note that the classical ER percolation is equivalent to our $k=2$
case, and the ER order parameter (relative number of edges) is
identical to $\Psi$.
Also note that in general the size of the largest cluster could be
measured as the number of its $l$-cliques,
${\cal N}^*_{(l)}$, for $1\leq l \leq k$.
However, for simplicity we restrict ourselves to the two limiting cases
($N^* \equiv {\cal N}^*_{(1)}$ and ${\cal N}^* \equiv {\cal N}^*_{(k)}$)
defined above.

\begin{figure}[t!]
\centerline{\includegraphics[angle=0,width=0.85\columnwidth]{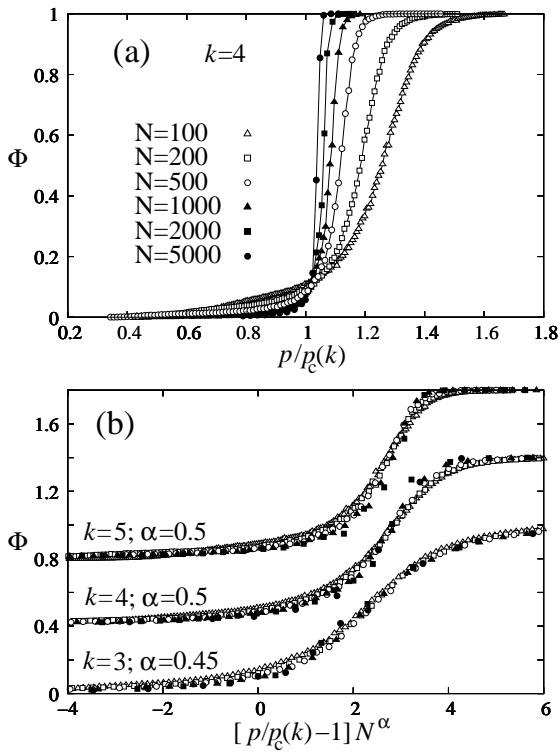}}
\caption{
Simulation results for the order parameter $\Phi$ averaged over several
runs (the statistical error is smaller than the symbol size).
(a) The convergence of $\Phi$ as a function of $p/\pc(k)$ to a step
function in the $N\to\infty$ limit is illustrated for $k=4$.
(b) The width
of the steps follows a power law, $\sim N^{-\alpha}$, as the steps
collapse onto a single curve if we stretch them out by $N^{\alpha}$
horizontally. The data for $k=4$ and 5 are shifted upward by 0.4 and
0.8, respectively, for clarity.
}
\label{fig_phi}
\end{figure}

\begin{figure}[t!]
\centerline{\includegraphics[angle=0,width=0.85\columnwidth]{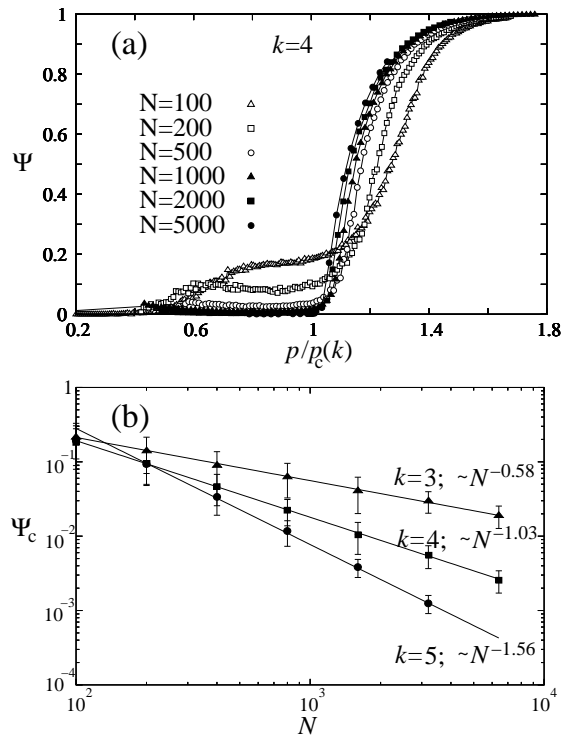}}
\caption{
The order parameter $\Psi$ for the same simulations as in Fig.\
\ref{fig_phi}.
(a) As illustrated for $k=4$, $\Psi$ as a function of $p/\pc(k)$
converges to a limit function (which is 0 for $p/\pc(k)<1$ and grows
continuously to 1 above $p/\pc(k)=1$) in the $N\to\infty$ limit. (b)
The order parameter at the threshold, $\Psi_{\rm c}$, scales as some
negative power of $N$, in good agreement with expression (\ref{Psi_c}).
}
\label{fig_psi}
\end{figure}

Our computer simulations indicate that the two order parameters behave
differently near the threshold probability. To illustrate this, in
Figs.\ \ref{fig_phi}a and \ref{fig_psi}a we plotted
$\Phi$ and $\Psi$, respectively, as a function of $p/\pc(k)$ for $k=4$
and for various system sizes ($N$), averaged over several runs.

The order parameter $\Phi$ for $k\geq 3$ converges to a step
function as $N\to\infty$. The fact that the step is located at
$p/\pc(k)=1$ is actually the numerical proof of the validity of our
theoretical prediction (\ref{pcrit}) for $\pc(k)$.
The width of the steps follows a power law,
$\sim N^{-\alpha}$, with some exponent $\alpha$.
Plotting $\Phi$ as a function of $[p/\pc(k)-1]N^{\alpha}$, \ie,
stretching out the horizontal scale by $N^{\alpha}$, the data
collapse onto a single curve. This is shown for $k=3$, 4, and 5 in
Fig.\ \ref{fig_phi}b. The exponent $\alpha$
seems to be around 0.5 for $k\geq 3$. Although for $k=3$ a slight
deviation form $\alpha=0.5$ has been obtained, we cannot distinguish that
from a possible logarithmic correction.

The order parameter $\Psi$ for $k\geq 2$, on the other hand, similarly
to the classical ER transition, converges to a
limit function, which is
0 for $p/\pc(k)<1$ and grows continuously from 0 to 1 if we increase
$p/\pc(k)$ from 1 to $\infty$.

One of the most fundamental results in random graph theory concerns the
behavior of the largest component at the percolation threshold, where
it becomes a giant (infinitely large) component in the
$N\to\infty$ limit. Erd\H{o}s and R\'enyi showed
\cite{e-r}
that for the random graphs they introduced, the size of the largest
component ${\cal N}^*$ (measured as the number of its edges) at
$p=\pc \equiv 1/N$ diverges with the system size as
$N^{2/3}$, or equivalently, the order parameter $\Psi$ scales as
$N^{-1/3}$. Since the giant component at the threshold has a tree-like
structure, its number of vertices, $N^*$, also diverges as $N^{2/3}$.
We shall show that similar scaling behavior can be obtained
for $k$-clique percolation at the threshold probability $\pc(k)$.

If we assume, that the $k$-clique adjacency graph is like
an ER graph
\cite{assumption},
then at the threshold the size of its giant component
${\cal N}_{\rm c}^*$ scales as ${\cal N}_{\rm c}^{2/3}$.
The subscript ``c'' throughout this Letter indicates that
the system is at the
percolation threshold (or critical point). Plugging $p=\pc$ from
Expression (\ref{pcrit}) into Eq.\ (\ref{Nk}) and omitting the
$N$-independent factors we get the scaling
\be
{\cal N}_{\rm c} \sim N^{k/2}
\label{Nkc}
\ee
for the total number of $k$-cliques.
Thus, the size of the giant component ${\cal N}_{\rm c}^*$ is expected
to scale as
${\cal N}_{\rm c}^{2/3} \sim N^{k/3}$
and the order parameter $\Psi_{\rm c}$ as
${\cal N}_{\rm c}^{2/3}/{\cal N}_{\rm c} \sim N^{-k/6}$.

This is valid, however, only if $k\leq 3$. The reason for the
breakdown of the above scaling is that for $k>3$ it predicts that the
number of $k$-cliques of the giant $k$-clique percolation cluster, \ie, the
number of vertices of the giant component in the $k$-clique adjacency
graph,
${\cal N}_{\rm c}^{2/3} \sim N^{k/3}$,
grows faster than $N$. On the other hand, in analogy with the structure
of the giant component of the classical ER problem, we expect that the
giant component in the adjacency graph also has a tree-like structure
at the threshold, with very few loops. As a consequence, almost every
vertex of the adjacency graph corresponds to a vertex of the
original graph. Thus, in the adjacency graph the giant component should
not grow faster than
$N$ at the threshold. Therefore, for $k>3$ we expect that
${\cal N}_{\rm c}^* \sim N$, and using Eq.\ (\ref{Nkc}),
$\Psi_{\rm c} = {\cal N}_{\rm c}^*/{\cal N}_{\rm c} \sim N^{1-k/2}$.
In summary:
\be
\Psi_{\rm c} \sim
\left\{ \begin{array}{ll}
 N^{-k/6}  & \textrm{ for } k \leq 3\\
 N^{1-k/2} & \textrm{ for } k \geq 3
\end{array} \right.
.
\label{Psi_c}
\ee
We have also determined the scaling of $\Psi_{\rm c}$ at $\pc$ as a
function of $N$ numerically, and the results are in good agreement
with the above heuristic arguments, as shown in Fig.\ \ref{fig_psi}b.

Finally, we discuss the relevance of our approach to community finding
\cite{domany-prl,gn-pnas,zhou,newman-fast,al-parisi,huberman,spektral}.
If we start rolling a $k$-clique template from a highly connected part of
a network we can proceed and find all vertices that can be reached
from the initial $k$-clique. Such a $k$-clique percolation cluster can
be identified as a community, because of the many (at least $k-1$)
links of any of its vertices to the other vertices of this cluster. The
links are organized into complete subgraphs ($k$-cliques), which is
also a characteristics of most communities (just think of human relations).
With different values of $k$ we can identify communities of different
strength (or cohesiveness).
Our $k$-clique percolation clusters also satisfy a number of basic requirements
(local; density based; not too restrictive; have no cut-node; allow overlaps)
that are expected from a community definition, but are not satisfied
simultaneously by any other existing definition in the literature
\cite{clique-overlap,kosub}.
Although using $k$-cliques might seem to be a very strict constraint on
the community definition, we note that relaxing this constraint (\eg,
by allowing incomplete $k$-cliques) is practically equivalent to
lowering the value of $k$.

The sharp percolation transition (step in $\Phi$) of the ER graphs
provides the theoretical basis for the applicability of our community
definition to real networks. This is because if the network was
completely random, only very few and small clusters would be expected
for any $k$ at which the network is below the transition point.
However, if large clusters do appear, they must correspond to locally
dense structures, \ie, real communities. Moreover, since these
communities are locally above the percolation threshold, their
identification is immune to random removal of edges as long as their
edge density remains above the threshold.

The most important aspect of such a method is that naturally, a single
vertex can be part of several communities
\cite{Pallasubm},
as illustrated in Fig.\ \ref{fig_sketch} (right) by the half black,
half dark gray vertex.
In terms of a person, he/she can
belong to a number of groups (of highly connected people) in such
a way that no two groups share a ($k-1$)-clique (there are no
$k-1$ people in any two groups who would all know each
other and, therefore, would allow a $k$-clique template to roll through).
Thus, each vertex can belong to a number of individually
identifiable communities and, in turn, each community can have a
large number of contacts with other communities,
just as it happens in most realistic situations
(see, \eg, Ref.\
\cite{scott-book}).
This is very much in contrast with the divisive and agglomerative
methods, which force each vertex to belong to only one community and be
separated from the others, leading to the loss of many of the
communities of the network.

The approach presented in this Letter allows a number of
generalizations (\eg, $k$-cliques connected through
($k-l$)-cliques, $k$-cliques with weighted edges, etc.) and opens
new directions in the study of network structures made of highly
interconnected parts including communities overlapping in various
non-trivial ways. As an important biological example, we have
successfully applied our method to the identification of protein
communities in
the protein-protein interaction network of yeast, which has allowed us
to make predictions for the yet unknown function of some proteins
\cite{Pallasubm}.

\begin{acknowledgments}
This work has been supported in part by
the Hungarian Science Foundation (OTKA), grant Nos. F047203 and T034995.
\end{acknowledgments}

\vspace{-6pt}							


\begin{thebibliography}{}
\bibitem{assumption}
This assumption is an approximation since the adjacency graph is
weakly assortative.
\end{thebibliography}
\end{document}